\documentstyle[epsfig,12pt]{article}
\begin{document}
\baselineskip=15pt

\begin{center}

%%%%%%%%%%%%%%%%%%%%%%%%%%%%%%%%%%%%%%%%%%%%%%%%%%%%%%%%%%%%%%%%%%%%%%%%%%%%%%%
{\Huge \bf --------------------------}

{\tiny 
Ochanomizu University Workshop; June 14,  2001
}

\vskip .5cm

{\Huge \bf --------------------------}
%%%%%%%%%%%%%%%%%%%%%%%%%%%%%%%%%%%%%%%%%%%%%%%%%%%%%%%%%%%%%%%%%%%%%%%%%%%%%%%

\end{center}
\vskip 1cm

\begin{center}
{\Large\bf

The Infrared Boundary of Perturbative QCD}

\vskip .8cm

Victor {\sc ELIAS $^{a,}$}
\footnote{Electronic address: {\tt velias@uwo.ca}}
%,
%Name 2 {\sc Lastname 2 $^{b,}$}
%\footnote{Electronic address: {\tt author2@...}} 
%, \\
%Name 3 {\sc Lastname 3 $^{c,}$}
%\footnote{Electronic address : {\tt author2@...}}\\

\vskip .8cm
{\small \it
$^a$ 
Department of Applied Mathematics, The University of Western Ontario,  London, Ontario  N6A 5B7 CANADA
\\
%$^b$ 
%Department of Author 2, University of Author 2, Address of Author 2 
%\\
%$^c$ 
%Department of Author 3, University of Author 3, Address of Author 3
}
\end{center}

\vskip 1cm	

%%%%%%%%%%%%%%%%%%%%%%%%%%%%%%%%%%%%%%%%%%%%%%%%%%%%%%%%%%%%%%%%%%%%%%%%%%%%%%%
\begin{center}
{\Large \bf ------------------------------------------------------------}
\end{center}
\vskip 1cm	

\begin{center}
{\bf  Abstract}\\
\end{center}
{\small \sl

Evidence is reviewed suggesting that QCD remains a perturbative theory with a 
(relatively) small coupling constant down to a distinct infrared
boundary on perturbative physics, a boundary corresponding to the
momentum scale associated with a $\beta$-function pole.

}

\sl

\newpage
%%%%%%%%%%%%%%%%%%%%%%%%%%%%%%%%%%%%%%%%%%%%%%%%%%%%%%%%%%%%%%%%%%%%%%%%%%%%%%%

\setcounter{footnote}{0}

\section{Supersymmetric Gluodynamics}

Our information about the coupling constants characterising known
interactions is necessarily perturbative.  Many QCD calculations, for
example, have been performed in the $\overline{MS}$ scheme in which the
$\beta$-function for the couplant $x \equiv \alpha_s (\mu)/\pi$ is known
to three subleading orders \cite{1}:
\begin{equation}%1
\mu^2 \frac{dx}{d\mu^2} \equiv \beta(x) = -\beta_0 x^2 - \beta_1 x^3 -
\beta_2 x^4 - \beta_3 x^5 - \beta_4 x^6 ...
\end{equation}
\begin{eqnarray}%2
\beta_0 & = & 11/4 - n_f / 6 \nonumber\\
\beta_1 & = & 51/8 - 19n_f / 24 \nonumber \\
\beta_2 & = & 2857/128 - 5033 n_f/1152 + 325 n_f^2 / 3456 \nonumber \\
\beta_3 & = & 114.23033 - 27.133944 n_f + 1.582379 n_f^2 + 5.856696
\cdot 10^{-3} n_f^3
\end{eqnarray}
Thus, we can determine with high accuracy how the QCD couplant evolves
with $\mu$ in perturbatively accessible processes.  However, the
behaviour of the couplant in the infrared region, the region where
successive terms in the series (1) become comparable in magnitude, is
not at all clear from the truncated perturbative series.

There is value in having theories in which the $\beta$-function {\it is}
exactly known.  One such theory is $N = 1$ supersymmetric $SU(N_c)$
Yang-Mills theory, a theory of gluons and gluinos often denoted as
``supersymmetric gluodynamics.''  The $\beta$-function for this theory
has been computed to all orders by Novikov, Shifman, Vainshtein, and
Zakharov (NSVZ) using instanton calculus methods \cite{2};  it is also
obtainable algebraically by requiring that the supersymmetric anomaly
multiplet characterising the theory satisfy the Adler-Bardeen theorem
\cite{3,4}:
\begin{eqnarray}%3
\mu^2 \frac{dx}{d\mu^2} & \equiv & \beta^{NSVZ}(x)= - \frac{3N_c x^2}{4}
\left[ \frac{1}{1-N_c x/2} \right], \nonumber\\
x & \equiv & g^2 / 4\pi^2
\end{eqnarray}
The pole characterising the $\beta$-function (3) necessarily implies the
existence of an infrared boundary for the asymptotically-free phase of
the couplant \cite{5}.  Suppose we have some sufficiently small initial
value $x_0$ $[x_0 < 2/N_c]$ occurring at momentum scale $\mu_0$.  We
see that the $\beta$-function is negative, implying that $x(\mu)$ grows
as $\mu$ decreases until $x = 2/N_c$, at which point $\beta(x)$ becomes
singular.  Alternatively $\mu$ can be regarded as a function of $x$,
with initial value $\mu_0$ occurring at $x_0$ and with an extremum
occurring when $x = 2/N_c$. This extremum is necessarily a minimum, a
critical value $\mu_c$ that represents the lower bound on the domain of
$x(\mu)$.  In short, supersymmetric gluodynamics is a field theory
characterised by a real perturbative interaction coupling constant only when $\mu >
\mu_c$.  Not only is $\mu_c$ the infrared boundary of this theory, but
this boundary is characterised by a {\it finite} value of the
couplant:  $x(\mu_c) = 2/N_c$.

The behaviour described above is highly suggestive of the qualitative picture we have of
the strong interactions, whose perturbative character as a theory of quarks and gluons abruptly
ends at near-GeV hadronic mass scales. At such mass scales, strong interactions are described by
some effective-Lagrangian phenomenological theory distinct from perturbative QCD, a theory in 
which fundamental fields are no longer quarks and gluons, but hadrons.

\section{Non-Supersymmetric Gluodynamics}

For a function underlying an incompletely known perturbative series, such as the
QCD $\beta$-function (1), one of the few available techniques for extracting information 
about the existence of poles or zeros of that function is to examine Pade approximants 
constructed entirely from the known terms of the series.  Such approximants may exhibit 
a positive zero which precedes any positive poles, behaviour indicative of a nonzero 
infrared-stable fixed point. Alternatively, a leading pole, {\it e.g.},  a positive pole 
which precedes any positive zeros in the approximant, is indicative of a $\beta$-function 
analogous to (2), a function characterised by an infrared boundary to perturbative dynamics 
in its asymptotically free phase.  While one particular Pade approximant
may exhibit a spurious pole (usually denoted as a ``defect pole''), the occurrence of 
leading poles in many different Pade approximants all constructed from the same truncated 
series is strong evidence that such a pole genuinely occurs within the function 
underlying that series \cite{6}. \footnote{A demonstration of how Pade approximant poles 
for a series that differs only infinitesimally from a geometric series are able to 
reproduce the geometric-series pole is presented in ref. [4].}

     For ($N_c = 3$) gluodynamics, this truncated series is given by eqs. (1) and (2) with 
$n_f = 0$.  The lowest approximants possessing both zeros and poles that are determined 
by the known and first unknown terms of the series (1) are
\begin{equation}%4
\beta^{[2|1]} (x) = -\frac{11}{4} x^2 \left[ \frac{1-2.7996x -
3.7475x^2}{1-5.1178x} \right]
\end{equation}
\begin{equation}%5
\beta^{[1|2]} (x) = -\frac{11}{4} x^2 \left[ \frac{1-5.9672x}{1-8.2854x+11.091x^2} \right]
\end{equation}
\begin{eqnarray}%6
\beta^{[3|1]} & = & - \frac{11}{4} x^2 \left[1+(2.31818 - 0.0087542\beta_4)x+(8.11648 - 0.020294\beta_4)x^2 \right. \nonumber\\
& + & \left. (41.5383-0.071053\beta_4)x^3\right] / \left[ 1-0.0087542\beta_4 x \right]
\end{eqnarray}  
\begin{eqnarray}%7
\beta^{[2|2]} (x) = - \frac{11}{4} x^2 \frac{ \left[ 1+(13.4026 -
0.027715\beta_4) x + (-22.9153+0.032788\beta_4) x^2 \right]}{
\left[1+(11.0844-0.027715\beta_4)x+(-56.7275+0.097036\beta_4) x^2
\right]}
\end{eqnarray}
\begin{eqnarray}%8
\beta^{[1|3]} & = & -\frac{11}{4} x^2 \left[1+(9.56239-
0.022220\beta_4)x\right] / \left[1+(7.24421-0.022220\beta_4)x \right.
\nonumber\\
& + & \left. (-24.9099+0.005151\beta_4)x^2+(-42.5901+0.060939\beta_4)x^3\right]
\end{eqnarray}  
The Maclaurin expansions of the first two approximants reproduce
the known terms $\beta_0 - \beta_3$ of the series (1) when $n_f=0$.  The
Maclaurin expansions of the final three approximants also reproduce the
unknown $\beta_4$ term of that series.  In (4) and (5), the first
positive pole [$x=0.195$ and $x=0.151$, respectively] is seen to precede
the first positive zero [$x=0.264$ and $x=0.168$].
Remarkably, this behaviour persists in (6), (7) and (8), regardless of
the value of the unknown coefficient $\beta_4$ in the $\beta$-function
series (1), as is demonstrated graphically in Figs. 3-5 of ref. \cite{7}.
In (6), for example, a positive pole occurs at $x=114.2/\beta_4$
provided $\beta_4$ is positive.  This pole is always smaller than any
positive zeros of the degree-3 numerator polynomial. Moreover, when
$\beta_4$ is negative, eq.(6) exhibits no positive zeros {\it or}
poles.  Eqs. (7) and (8), which have higher-degree denominator
polynomials, exhibit a positive pole for all values of $\beta_4$ which
is seen to always precede any numerator zeros for that same value of $\beta_4$.
Thus, for Pade approximants to the $\beta$-function capable of generating both
zeros and poles, a zero {\it never} precedes a pole for $N_c = 3$
gluodynamics, as would be expected if gluodynamics were to have an
infrared stable fixed point.  Instead, all such approximants point
strongly toward the existence of a $\beta$-function pole, as is known to
characterise supersymmetric gluodynamics in the NSVZ renormalization
scheme.

Such behaviour also characterises QCD in the 'tHooft ($N_c \rightarrow
\infty$) limit, a ``gluodynamics'' for any finite choice of $n_f$.  This
is evident from the five approximants analogous to (4-8) constructed
from the known and leading unknown coefficient in the $\beta$-function
\cite{1} describing the evolution of the (finite) couplant $\lambda
\equiv N_c \alpha_s (\mu) / 4 \pi$ as $N_c \rightarrow \infty \; (\alpha_s
\rightarrow 0)$:
\begin{equation}%9
\mu^2 \frac{d\lambda}{d\mu^2} = -\frac{11}{3} \lambda^2 - \frac{34}{3}
\lambda^3 - \frac{2857}{54} \lambda^4 - 315.49 \lambda^5 - \beta_4
\lambda^6 ...
\end{equation}
These five approximants are presented in the Appendix to ref. \cite{7}.
The pole/zero structure of these approximants is the same as that of
$N_c = 3$ gluodynamics -- a positive pole always precedes any positive
zeros occurring within leading approximant versions of the 
$\beta$-function (9).  These results strongly point to the existence of an
infrared boundary to gluodynamics as a perturbative theory, a boundary
which occurs at the value of $\mu$ associated with such poles.

\section{QCD}

In ref. \cite{7}, the dynamics described above for gluodynamics are
shown to apply to all approximant versions of the QCD $\beta$-function,
even if $n_f$ is as large as 5.  When $n_f \geq 6, \; \beta^{[2|1]}$ no longer
exhibits a pole, and poles are seen to precede zeros in $\beta^{[2|2]}$,
$\beta^{[2|1]}$ and $\beta^{[1|3]}$ only if $\beta_4$ is larger than
approximant-specific lower bounds.  Such results corroborate a lattice
study indicative of a similar $n_f$ threshold $(n_f = 7)$ for infrared-stable
fixed points to occur within QCD \cite{8}.

Pade approximant methods have been used successfully to predict unknown
QCD $\beta$-function terms.  For example, such methods, accompanied by
explicit knowledge of the ${\cal{O}}(n_f^3)$ term in $\beta_3$ led
Ellis, Karliner and Samuel \cite{9} to predict $\beta_3$ in (1):
\begin{equation}%10
\beta_3^{pred} = \frac{23,600 - 6400 n_f + 350 n_f^2 + 1.499 n_f^3}{256}.
\end{equation}
Comparison of this result with the exact result (2), as calculated
explicitly in \cite{1}, demonstrates the power of Pade approximant
methods.  Similar methods have been used to obtain a corresponding
prediction for the presently-unknown coefficient $\beta_4$ \cite{10}:
\begin{equation}%11
\beta_4^{pred} = \frac{759,000-219,000 n_f+20,500 n_f^2 - 49.8 n_f^3 -
1.84 n_f^4}{1024}.
\end{equation}

In attempting to extract information about the infrared boundary of
perturbative strong interaction physics, our interest is necessarily
directed toward the $n_f = 3$ case of (11), for which $\beta_4^{pred} = 278$.
The Pade approximant versions of the $n_f = 3$ QCD $\overline{MS}$
$\beta$-function whose Maclaurin expansions reproduce this estimate of
$\beta_4$ as well as $\beta_1 - \beta_3$, as given in (2), are \cite{11}
\begin{equation}%12
\beta^{[3|1]} (x) = -\frac{9}{4} x^2 \frac{\left[1-4.116x-6.006x^2-5.359x^3\right]}
{1-5.893x}
\end{equation}
\begin{equation}%13
\beta^{[2|2]} (x) = -\frac{9}{4} x^2 \frac{\left[1-5.498x-1.972x^2\right]}
{\left[1-7.276x+6.492x^2\right]}
\end{equation}
\begin{equation}%14
\beta^{[1|3]} (x) = -\frac{9}{4} x^2 \frac{\left[1-5.740x\right]}
{\left[1-7.517x+8.893x^2-3.190x^3\right]}
\end{equation}
All three approximants exhibit positive poles that precede any positive zeros. 
These poles $x_c$ respective occur at
\begin{equation}%15
x_c^{[3|1]} = 0.170, \; \; x_c^{[2|2]} = 0.160, \; \; x_c^{[1|3]} =
0.162
\end{equation}
and correspond to values of $\alpha_s$ between 0.50 and 0.53.

Both the agreement exhibited by these distinct approximants on the
magnitude of their poles, as well as the perturbatively small value of the
$\alpha_s$ corresponding to these poles, are quite striking.  The mass
scale $\mu_c$ associated with these approximant poles is easily
obtained from integrating the differential equation
\begin{equation}%16
\mu^2 \frac{d}{d\mu^2} x = \beta^{[N|M]} (x)
\end{equation}
from a physical initial value $x_0 (\mu_0)$ to the pole value
$x_c^{[N|M]}$:
\begin{equation}%17
\mu_c^{[N|M]} = \mu_0 \exp \left[ \frac{1}{2} \int_{x_0}^{x_c^{[N|M]}}
\frac{dx}{\beta^{[N|M]} (x)} \right].
\end{equation}
If we identify $\alpha_s (m_\tau) = 0.314 \pm 0.010$ \cite{12} and if we
require the couplant to devolve with four active flavours from $\mu =
m_\tau$ to a four-flavour threshold at 1.2 GeV (below which only three
active flavours contribute), we find that
\begin{equation}%18
\mu_c^{[3|1]} = 950 \pm 50 \; MeV, \; \; \mu_c^{[2|2]} = 990 \pm 50 \;
MeV, \; \; \mu_c^{[1|3]} = 980 \pm 50 \; MeV.
\end{equation}
Alternatively, we can utilise the (three-active flavour) estimate
$\alpha_s (m_\tau) = 0.33 \pm 0.02$ \cite{13} to find entirely via 
(12-14) and (17) that \cite{11}
\begin{equation}%19
\mu_c^{[3|1]} = 1.09 \pm 0.11 \; GeV, \; \; \mu_c^{[2|2]} =
1.14 \pm 0.11 \; GeV, \; \; \mu_c^{[1|3]} = 1.13 \pm 0.11 \; GeV.
\end{equation}
These estimates are all suggestive of the $\mu_c \simeq 4\pi f_\pi$
(=1.17 GeV) Georgi-Manohar boundary on effective-Lagrangian strong
interaction physics \cite{14}.  These results also show remarkable
consistency across the approximants (12-14) in the coordinates of the
infrared terminus ($\mu_c, x_c \equiv \alpha_s (\mu_c) / \pi$) of QCD as
a perturbative theory of quarks and gluons.

Of course, these estimates rely on the estimate (11) for $\beta_4$.
Because of the alternation of sign in this expression, the estimate
itself is considerably less likely to be accurate than the individual
estimates for polynomial coefficients of powers of $n_f$.  An alternative 
approach toward acquiring insight into the infrared boundary of QCD is 
{\it to assume} that $\alpha_s(\mu_c) = \pi/4$, the threshold value associated 
with dynamical chiral symmetry breaking \cite{15}.  Individual $[3|1]$, $[2|2]$, and
$[1|3]$ approximants to the $n_f = 3$ $\beta$-function when $\beta_4$ is
taken to be arbitrary are given by
\begin{eqnarray}%20
\beta^{[3|1]} (x) & = & -\frac{9}{4} x^2 \left[ 1+(1.7778-0.021174 \beta_4)x 
+ (4.4471-0.037642 \beta_4)x^2 \right. \nonumber\\ 
& + & \left. (20.990-0.094670\beta_4)x^3\right] / \left[1-0.021174\beta_4 x \right]
\end{eqnarray}
\begin{equation}%21
\beta^{[2|2]} (x) = -\frac{9}{4} x^2 \frac{\left[1+(7.19456-0.045604\beta_4)x+(-11.3292+0.033619\beta_4)x^2\right]}
{\left[1+(5.41678-0.045604\beta_4)x+(-25.4301+0.11469\beta_4)x^2\right]}
\end{equation}
\begin{eqnarray}%22
\beta^{[1|3]} (x) & = & -\frac{9}{4} x^2 \left[1+(5.80845-0.041491\beta_4)x \right] / \left[ 1+(4.03067-0.041491\beta_4)x \right. \nonumber\\
& + & \left. (-11.6367+0.073762 \beta_4)x^2 +(-18.3242+0.054377\beta_4)x^3 \right]
\end{eqnarray}
These approximants are respectively seen to develop a leading positive pole at $x=1/4$ ({\it i.e.} $\alpha_s = \pi/4$) for the following values
of $\beta_4$:
\begin{equation}%23
\beta_4^{[3|1]}=189, \; \; \beta_4^{[2|2]}=181, \; \; \beta_4^{[1|3]}=202,
\end{equation}
values surprisingly uniform across the three distinct approximants.  Upon substituting 
such values for $\beta_4$ into their corresponding approximants (20-22),
we can estimate via (17) concomitant values for the infrared-boundary
momentum scale at the $x = 1/4$ pole.  We set $x_c^{[N|M]} = 1/4$ in
(17) and utilize the extreme values of the 3-active-flavour range $0.33
\pm 0.02$ already quoted for $\alpha_s(m_\tau)$ in order to find that
\cite {11}
\begin{eqnarray*}
\alpha_s(m_\tau)=0.31: \mu_c^{[3|1]} = 775 \; MeV, \; \;
\mu_c^{[2|2]}=785 \; MeV, \; \; \mu_c^{[1|3]} = 778 \; MeV
\end{eqnarray*}
\begin{eqnarray*}
\alpha_s(m_\tau)=0.33: \mu_c^{[3|1]} = 863 \; MeV, \; \;
\mu_c^{[2|2]}=874 \; MeV, \; \; \mu_c^{[1|3]} = 867 \; MeV
\end{eqnarray*}   
\begin{eqnarray*}
\alpha_s(m_\tau)=0.35: \mu_c^{[3|1]} = 948 \; MeV, \; \;
\mu_c^{[2|2]}=960 \; MeV, \; \; \mu_c^{[1|3]} = 952 \; MeV.
\end{eqnarray*}
There is remarkable uniformity in these estimates obtained from distinct
approximants, suggesting that the infrared boundary is genuine, rather
than a single-approximant artefact.  A more precise estimate of the
boundary, however, requires more precise knowledge of $\alpha_s$ at
$\mu=m_\tau$.  We can conclude, however, that there is clear evidence from
Pade approximant methods for QCD to be a computationally-perturbative 
gauge theory of quarks and gluons right up to an ${\cal{O}}(1 \; GeV)$ 
boundary mass scale, an explicit lower bound on the domain of $\alpha_s(\mu)$
below which the description of the strong interactions must necessarily be quite
different.

\vskip 1cm
{\noindent \large\bf Acknowledgments:}\\
 
I am most grateful to Professor A. Sugamoto both for organising this workshop
on short notice and for providing us with gracious 
hospitality during our visit to Ochanomizu 
University. I am also grateful to the International Opportunity Fund of 
the Natural Sciences and Engineering Research Council of Canada for financial
support of this meeting.

%%%%%%%%%%%%%%%%%%%%%%%%%%%%%%%%%%%%%%%%%%%%%%%%%%%%%%%%%%%%%%%%%%%%%%%%%
%\begin{center}
%{\Large \bf ------------------------------------------------------------}
%\end{center}
%\begin{center}
%{\Large \bf Appendixes}
%\end{center}
%\begin{center}
%{\Large \bf ------------------------------------------------------------}
%\end{center}

%%%%%%%%%%%%%%%%%%%%%%%%%%%%%%%%%%%%%%%%%%%%%%%%%%%%%%%%%%%%%%%%%%%%%%%%%

%\appendix

\end{document}